\documentclass[aps,prl,showpacs,twocolumn,groupedaddress]{revtex4} 
\usepackage{amsmath}
\usepackage{graphicx}
\usepackage{bm}

\renewcommand{\Im}[1]{\ensuremath{{\rm Im}\left(#1\right)}}
\renewcommand{\Re}[1]{\ensuremath{{\rm Re}\left(#1\right)}}

\begin{document}

\title{Three-dimensional negative index of refraction at optical frequencies by coupling plasmonic waveguides}

\author{Ewold Verhagen}
 \email{verhagen@amolf.nl}
\author{Ren\'e de Waele}
\author{L. (Kobus) Kuipers}
\author{Albert Polman}
\affiliation{Center for Nanophotonics, FOM Institute for Atomic and Molecular Physics (AMOLF), Science Park 104, 1098 XG, Amsterdam, The Netherlands}

\date{\today}

\begin{abstract}
We identify a route towards achieving a negative index of refraction at optical frequencies based on coupling between plasmonic waveguides that support backwards waves. We show how modal symmetry can be exploited in metal-dielectric waveguide pairs to achieve negative refraction of both phase and energy. By properly controlling coupling between adjacent waveguides, a metamaterial consisting of a one-dimensional multilayer stack exhibiting an isotropic index of $-1$ can be achieved at a free-space wavelength of 400~nm. The general concepts developed here may inspire new low-loss metamaterial designs operating close to the metal plasma frequency.
\end{abstract}

\maketitle

Metamaterials allow control over the propagation of electromagnetic waves in ways beyond those provided by naturally occurring materials. Since the prediction that a material with a negative index of refraction can be used to construct a perfect lens~\cite{pendry2000}, and spurred on by the paradigm of transformation optics~\cite{pendry2006leonhardt2006}, a quest for media that possess a negative index has resulted in a wide range of metamaterials operating in different frequency regimes~\cite{soukoulis2007}. Many designs of such `left-handed' materials are based on subwavelength metallic resonant scattering elements that exhibit a simultaneously negative electric and magnetic response to light. These geometries can be related to archetypical structures such as split rings at low frequencies, where $\varepsilon_\mathrm{m} \ll-\varepsilon_\mathrm{d}$ ($\varepsilon_\mathrm{m}$ and $\varepsilon_\mathrm{d}$ are the dielectric constants of the metal and its dielectric surrounding, respectively)~\cite{soukoulis2007}. However, using this route it has proven to be very difficult to achieve a negative index with acceptable losses at visible frequencies, i.e., close to the metal plasma frequency $\omega_\mathrm{p}$~\cite{merlin2009,dolling2007,xiao2009}. Moreover, most designs only show the desired behavior for a limited range of angles. These limitations are less severe in indefinite materials, which rely on hyperbolic dispersion to achieve negative refraction of energy~\cite{smith2004a,wangberg2006,fan2006a,yao2008}. Because the phase index in these materials is positive, indefinite media however do not possess all the same properties as left-handed media. In a left-handed medium the direction of energy flow is opposite to that of the wavevector ($\mathbf{k}\cdot\mathbf{S}<0$)~\cite{veselago2006,foteinopoulou2003a}. An ideal negative index metamaterial has a phase index that is independent of angle, which results in the wavevector $\mathbf{k}$ and Poynting vector $\mathbf{S}$ being entirely antiparallel. Both phase and energy are then refracted negatively at an interface of a positive index medium with the metamaterial.

In this work, we propose a strategy to design metamaterials that have a three-dimensional negative index of refraction at visible frequencies.
The working principle relies on coupling between plasmonic waveguides that are arranged in a stacked geometry.
We use the optical properties of metals close to the plasma frequency, where $-\varepsilon_\mathrm{d}<\varepsilon_\mathrm{m}<0$, which allow the constituting waveguides to support backwards surface plasmon polariton (SPP) waves~\cite{tournois1997,shvets2003,alu2006a,shin2006,stockman2007,dionne2008}.
We show that the symmetry of the waveguide modes is crucial to the realization of negative refraction in the metamaterial. By exploiting symmetric modes in arrays of plasmonic waveguide pairs, the coupling between adjacent waveguides can be tailored to achieve negative refraction of both wavevector and energy. The design strategy presented in this manuscript allows the attainment of an isotropic negative index of refraction of $-1$ for TM-polarized light in a one-dimensional array of plasmonic slab waveguides. The material operates at visible frequencies with comparatively small absorption losses.

\begin{figure}[]
\includegraphics[width=6.4cm]{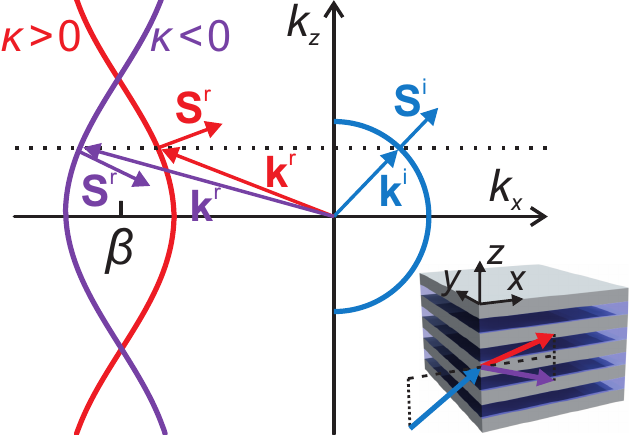}
\caption{(color) Wavevector diagram showing isofrequency contours of a homogeneous positive index medium (blue) and left-handed waveguide array with $\kappa>0$ (red) or $\kappa<0$ (purple).} \label{Fig1}
\end{figure}

Wave propagation in an array of coupled waveguides can be described by coupled mode theory in the weakly coupled paraxial limit~\cite{eisenberg2000christodoulides2003,fan2006a}. We consider a material comprised of slab waveguides aligned along the $xy$-plane and stacked in the $z$ direction (as sketched in the inset of Fig.~\ref{Fig1}). In the $xy$-plane, parallel to the slabs, the wavevector is independent of angle and its magnitude $\beta'$ will be slightly altered with respect to the mode wavevector $\beta$ in a single isolated waveguide due to the presence of the adjacent waveguides. The angle at which a beam is refracted that is incident in the $xy$-plane from a homogeneous medium onto an interface parallel to the $yz$-plane can therefore be found from Snell's law, using the refractive index $\beta' c/\omega$. For the material to be left-handed and refraction of both wavevector and energy to be negative in the $xy$-plane, the mode wavevector $\beta$ of the constituent waveguides must therefore be negative.

In the $xz$-plane, propagation and refraction is governed by waveguide coupling. Taking only nearest-neighbor interactions into account, the amplitudes in adjacent waveguides are related through~\cite{eisenberg2000christodoulides2003} 
\begin{equation}
\partial a_j / \partial x=i \beta a_j + i \kappa \left( a_{j-1} + a_{j+1} \right),
\label{eqdifferential}
\end{equation}
where $a_j$ is the complex amplitude of the mode in the $j^\mathrm{th}$ waveguide and $\kappa$ a coupling constant. Solving Eq.~(\ref{eqdifferential}) yields a simple relationship between the wavevector components parallel ($k_x$) and normal ($k_z$) to the waveguides:
\begin{equation}
k_x=\beta+2\kappa \cos \left(k_z d \right),
\label{eqcurve}
\end{equation}
where $d$ is the center-to-center distance between adjacent waveguides. This relationship is sketched in a wavevector diagram in Fig.~\ref{Fig1} for different signs of the coupling constant $\kappa$ and taking $\beta<0$.
Also indicated is the isofrequency contour for a homogeneous positive index medium (blue curve). Since the problem of interest is that of refraction of a plane wave incident in a homogeneous medium from $x=-\infty$ (see inset of Fig.~\ref{Fig1}), only curves are drawn for which $S_x>0$. By definition, the group velocity (and in the absence of loss also $\mathbf{S}$) is oriented normal to an isofrequency contour. Given a particular incident wavevector $\mathbf{k}^\mathrm{i}$, the refracted wavevector $\mathbf{k}^\mathrm{r}$ can be derived from the wavevector diagram by conserving the wavevector component $k_z$ (up to an integer times $2\pi/d$). We can see from Fig.~\ref{Fig1} that $\mathbf{k}^\mathrm{r}\cdot\mathbf{S}^\mathrm{r}<0$, which results from the fact that $\beta<0$. This material can therefore rightly be called left-handed, and wavefronts will thus be refracted negatively in the $xz$-plane as well as in the $xy$-plane. Importantly, however, whether the refraction of energy, given by the direction of $\mathbf{S}^\mathrm{r}$, is positive or negative is determined by the sign of the coupling constant $\kappa$. If $\kappa<0$, refraction of energy is negative (i.e., $S_z^\mathrm{i}S_z^\mathrm{r}<0$), whereas it is positive if $\kappa>0$. Three-dimensional negative refraction of both phase and energy can therefore only be realized when both $\beta$ and $\kappa$ are negative.

\begin{figure}[]
\includegraphics[width=7.8cm]{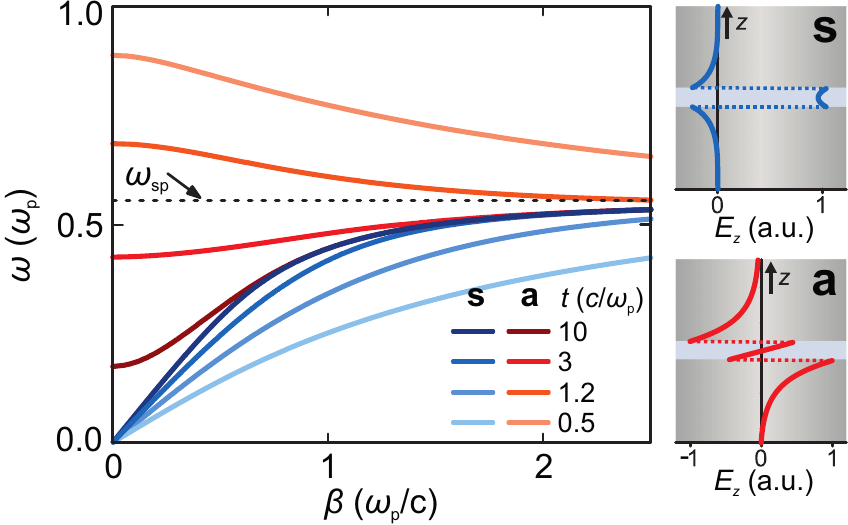}
\caption{(color) Dispersion in MDM waveguides for varying dielectric thickness $t$, showing the appearance of backwards SPPs for small $t$. On the right, characteristic $E_z$ profiles of the symmetric (\textbf{s}) and antisymmetric (\textbf{a}) SPP modes are displayed.} \label{Fig2}
\end{figure}

Guided modes with $\beta<0$ can be called `backwards' since energy and phase are counterpropagating~\cite{veselago2006}. They can exist in plasmonic waveguides in which more energy is guided in metal than in dielectric, since inside the metal $\mathbf{S}$ is oriented opposite to that in the dielectric. The antisymmetric SPP mode guided by a thin dielectric slab clad with metal, termed a metal-dielectric-metal (MDM) waveguide, can acquire a backwards character under certain conditions, as pointed out by several authors~\cite{tournois1997,shvets2003,alu2006a,shin2006,stockman2007,dionne2008}. Figure~\ref{Fig2} depicts the dispersion curves of the propagating symmetric and antisymmetric SPP modes in a MDM waveguide for various dielectric thicknesses $t$, calculated as described in~\cite{dionne2008} using $\varepsilon_\mathrm{m}=1-\omega_\mathrm{p}^2/\omega^2$ and $\varepsilon_\mathrm{d}=2.25$.
For simplicity, we neglect absorption here, but as we will see the same physics applies once it is included. The symmetry of the transverse electric field assigns the mode symmetry. The antisymmetric mode acquires a negative group velocity $d\omega/d\beta$ for thicknesses $t<\left(\pi c/2\omega_\mathrm{p}\right)\sqrt{\left(1+\varepsilon_\mathrm{d}\right)/\varepsilon_\mathrm{d}}$, when its cutoff frequency $\omega_\mathrm{c}=\left(2c/\sqrt{\varepsilon_\mathrm{d}}t\right)\tan^{-1}\sqrt{-\varepsilon_\mathrm{m}/\varepsilon_\mathrm{d}}$ becomes larger than the surface plasmon resonance frequency $\omega_\mathrm{sp}=\omega_\mathrm{p}/\sqrt{1+\varepsilon_\mathrm{d}}$. In the absence of absorption, this directly means that energy and phase are counterpropagating. Because for $\omega_\mathrm{sp}<\omega<\omega_\mathrm{p}$ the antisymmetric mode is the only propagating mode, the waveguide can be described by the effective index $\beta c/\omega$, where we have to use the `mirrored' branch with $\beta<0$ when we assume energy propagates forwards~\cite{dionne2008}. This can lead to negative refraction of light into such a waveguide~\cite{dionne2008,shin2006,lezec2007}. This effect is however purely two-dimensional, restricted to the plane of the waveguide.

\begin{figure}[]
\includegraphics[width=5cm]{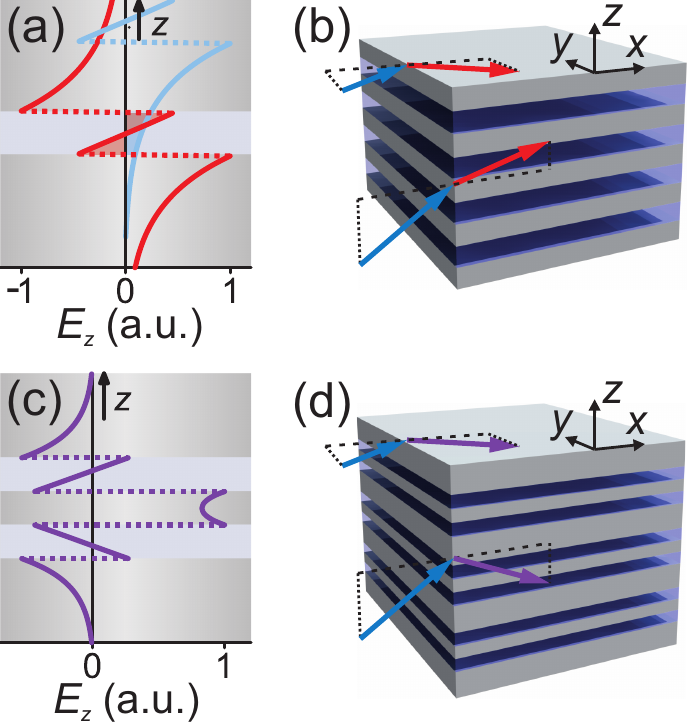}
\caption{(color) (a) $E_z\left(z\right)$ profile of the antisymmetric mode in a MDM waveguide (red), together with $E_z\left(z-d\right)$ (blue). (b) Refraction into an array of MDM waveguides. Arrows depict the direction of energy flow. (c) $E_z$ profile of the backwards symmetric mode in a pair of MDM waveguides. (d) Refraction into an array of strongly coupled MDM waveguide pairs.} \label{Fig3}
\end{figure}

Now we must ensure that these modes, when coupled in a three-dimensional array, lead to a negative coupling constant $\kappa$ to allow negative refraction of energy. The coupling constant can be approximated from the modal fields in an isolated waveguide by~\cite{marcuse}
\begin{equation}
\kappa = \omega \varepsilon_0 \frac{\left(\varepsilon_\mathrm{d}-\varepsilon_\mathrm{m}\right) \int_\mathrm{d} \mathbf{E}^{\dag}\left(z\right) \cdot \mathbf{E}\left(z-d\right) dz}{\int_{-\infty}^{\infty} \mathbf{\hat{x}} \cdot \left( \mathbf{E}^{\dag} \times \mathbf{H} + \mathbf{E} \times \mathbf{H}^{\dag} \right) dz}.
\label{eqkappa}
\end{equation}
The use of the adjoint fields $\mathbf{E}^{\dag}$ and $\mathbf{H}^{\dag}$ (obtained by substituting $-\beta$ for $\beta$ and $-\omega$ for $\omega$) is important for the correct treatment of absorbing media~\cite{supplementarynimcoupling,marcuse}. The integral in the numerator of Eq.~(\ref{eqkappa}) is performed over the dielectric region of the waveguide only~\cite{supplementarynimcoupling}. The sign of this mode overlap integral determines the sign of $\kappa$. Considering the mode profile $E_z\left(z\right)$ and the displaced field $E_z\left(z-d\right)$ (shown in Fig.~\ref{Fig3}(a)) of the antisymmetric mode in a MDM waveguide,
we see that the overlap integral $\int_\mathrm{d} E_z^{\dag}\left(z\right) E_z\left(z-d\right) dz$ is positive. This is the result of the fact that (1) $E_z$ switches sign across a metal-dielectric interface and (2) the mode is antisymmetric. A similar reasoning proves that $\int_\mathrm{d} E_x^{\dag}\left(z\right) E_x\left(z-d\right) dz$ is positive as well~\cite{supplementarynimcoupling}. As a result, a stack of equally spaced MDM waveguides as depicted in Fig.~\ref{Fig3}(b) will in fact have $\kappa>0$ and therefore exhibit positive refraction of energy in the $xz$-plane. It therefore does not constitute an ideal negative index material allowing negative refraction of both phase and energy in all directions.

As outlined above, the reason that $\kappa$ is positive is related to the antisymmetric nature of the mode in a MDM waveguide. To achieve negative refraction, a backwards mode with a symmetric nature is needed instead. This exists for example in a pair of strongly coupled MDM waveguides, in which eigenmodes resemble even and odd superpositions of the antisymmetric modes in the individual waveguides. The odd superposition has a mode wavevector $\beta$ that is slightly smaller than that of a single waveguide and also negative~\cite{footnotenimcoupling2}. Most importantly, this mode has a symmetric field profile (as shown in Fig.~\ref{Fig3}(c)), which causes the coupling constant $\kappa$ in Eq.~(\ref{eqkappa}) to be negative for coupling between adjacent waveguide pairs. An array of weakly coupled pairs of strongly coupled MDM waveguides, i.e., a metal-dielectric multilayer stack with alternating thick and thin metal layers as depicted in Fig.~\ref{Fig3}(d), will therefore exhibit negative refraction of both energy and wavevector in the $xz$-plane. An important second advantage to using the symmetric mode in a pair of MDM waveguides is that it can be excited by light incident along the $x$ axis from outside the metamaterial, whereas the antisymmetric mode has vanishing mode overlap with such a wave.

\begin{figure}[]
\includegraphics[width=8.6cm]{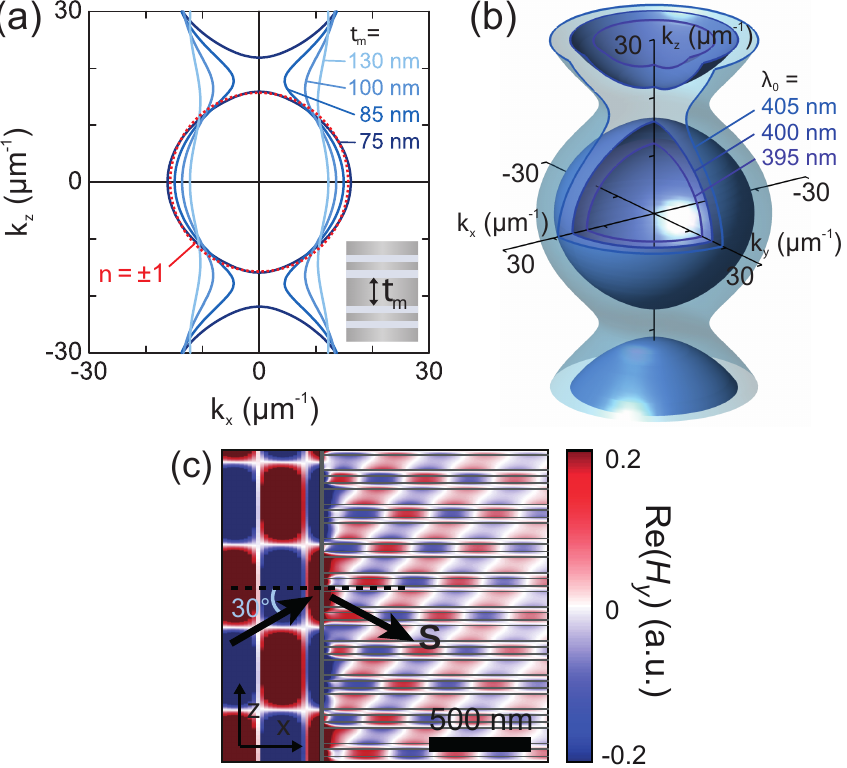}
\caption{(color) (a) Calculated wavevector diagrams for varying metal thickness between waveguide pairs, for the parameters described in the text. The red dotted curve shows the isofrequency contour for a homogeneous medium with $n=\pm1$. (b) Three-dimensional wavevector diagram showing isofrequency surfaces for three different values of $\lambda_0$, to show that $n$ is isotropic and negative. (c) Snapshot of the $H_y$ field distribution showing refraction of light from air into the metamaterial extending to the right of the vertical line, as calculated with FDTD. The arrows show the unit-cell-averaged direction of the Poynting vector.} \label{Fig4}
\end{figure}

An ideal negative index metamaterial has an isotropic phase index, which results in a circular isofrequency contour. The curvature of the isofrequency contour in Fig.~\ref{Fig1} is related to the coupling strength $\kappa$, which can be controlled by varying the thickness $t_\mathrm{m}$ of the metal layer separating adjacent pairs of waveguides. Figure~\ref{Fig4}(a) shows isofrequency contours for TM polarization for varying $t_\mathrm{m}$, calculated using a transfer matrix formalism.
The free-space wavelength $\lambda_0$ is 400~nm. We use the real part of the dielectric constant of Ag ($\varepsilon_\mathrm{m}=-4.43$~\cite{johnson1972}). We take $\sqrt{\varepsilon_\mathrm{d}}=3.2$, comparable to TiO$_2$ at this wavelength. A large dielectric index ensures $\omega_\mathrm{sp}$ is relatively small, extending the operating frequency window to a large part of the visible spectrum. The thicknesses of both dielectric layers and of the thin Ag film are 28 and 35~nm, respectively. Reducing $t_\mathrm{m}$ increases the coupling strength, resulting in a stronger modulation of $k_x$. Interestingly, the isofrequency contours evolve from the cosinusoidal behavior described in the weak coupling limit by Eq.~\ref{eqcurve} to almost fully circular for $t_\mathrm{m}=75$~nm. Since the structure is isotropic in the $xy$-plane, the isofrequency surface plotted in a three-dimensional wavevector diagram now resembles a sphere, as shown in Fig.~\ref{Fig4}(b). Wave propagation in the metamaterial can thus be described by an isotropic negative index of refraction $n$ that determines the refraction angle of both phase and group velocity. The magnitude of $n$ at a given frequency can be controlled most effectively by changing $\beta$ through varying the dielectric thickness (see Fig.~\ref{Fig2}). In this case, it has been chosen to result in $n=-1$. In Fig.~\ref{Fig4}(b), we have also indicated the isofrequency surfaces for two slightly different frequencies. The wavevector magnitude decreases for increasing frequency, confirming that the Poynting vector is oriented inwards and the material is left-handed.

It is interesting to note that for strong enough coupling, even propagation normal to the interfaces (along $\mathbf{\hat{z}}$) is allowed. The geometry and physical mechanism is different than for metal-dielectric multilayer stacks in the hyperbolic dispersion regime~\cite{ramakrishna2003}. 
The existence of transparency windows in metal-dielectric multilayers is well known, and can be attributed to resonantly coupled Fabry-P\'erot cavities~\cite{scalora1998}. In the currently proposed system, the fundamental harmonic of the Bloch wave propagating along $\mathbf{\hat{z}}$ can be described by a negative index. Of course, the fact that the wave comprises many Bloch harmonics has implications for the application of this metamaterial. For example, when used to construct a perfect lens, the largest magnitude of the wavevector component $k_z$ that can be transferred by the lens is $\pi/d$, i.e., the resolution along $\mathbf{\hat{z}}$ is limited by the dimension of a single waveguide. This can however be designed to be significantly smaller than $\lambda_0/2$.

When introducing realistic absorption losses in the previous example ($\varepsilon_\mathrm{m}=-4.43+0.21i$~\cite{johnson1972}), $\Re{n}$ is not notably affected~\cite{supplementarynimcoupling}. The figure of merit, defined as $\Re{\left|\mathbf{k}\right|}/\Im{\left|\mathbf{k}\right|}$, varies from 14.7 along $\mathbf{\hat{x}}$ to 8.7 along $\mathbf{\hat{z}}$. Considering the fact that the operating wavelength is only 400~nm in this metallodielectric metamaterial, absorption is thus comparatively small~\cite{dolling2007,valentine2008,xiao2009}, and not strongly dependent on the propagation direction.

To illustrate negative refraction of light into the metamaterial, Fig.~\ref{Fig4}(c) shows the result of a finite-difference time-domain simulation in which the interface of the metamaterial truncated along a plane parallel to $yz$ is illuminated with a TM-polarized plane wave incident under a $30^{\circ}$ angle from air. The material parameters are the same as before, and include absorption in Ag by locally fitting the dielectric constant to a Drude model with dissipation. The arrows indicate the unit-cell averaged direction of $\mathbf{S}$, showing that not only the wavefronts, but indeed also the energy is refracted negatively, with $n=-1$. The fraction of energy coupled to the refracted wave in the metamaterial is 2\%. However, in this example impedance matching was not optimized. Simulations have shown that coupling efficiencies up to 20\% are readily available by tuning the layer thicknesses~\cite{supplementarynimcoupling}. Moreover, structuring the interface may further improve impedance matching. Further studies should identify the limits and importance of impedance matching in a practical application, as well as derive values of effective permittivity and permeability for this metamaterial.

In conclusion, we have identified a route towards the construction of (left-handed) negative index metamaterials using plasmonic waveguide coupling. We underlined the role of backwards SPP modes and modal symmetry, and showed that the coupling can be controlled to achieve negative refraction of both energy and wavevector. A carefully designed one-dimensional stack of metallic and dielectric layers with can be characterized by a single three-dimensional isotropic negative index of refraction of $n=-1$ at a wavelength of 400~nm. In this respect, it differs strongly from photonic crystal slabs exhibiting negative index behavior~\cite{notomi2000,foteinopoulou2003}, which are essentially two-dimensional and in which backward wave behavior relies fully on Bragg diffraction. The structure proposed here only transmits TM-polarized light, but polarization-independent behavior could be found in designs that exploit alternative waveguide geometries~\cite{shvets2003,burgos2010}. Using Ag, the operating frequency of these metamaterials extends from the visible to the near-UV regime, with figures of merit of the order of 10. The same mechanisms can be applied throughout the electromagnetic spectrum, using conductive oxides in the near-infrared~\cite{west2010}, SiC in the mid-infrared~\cite{shvets2003}, (doped) semiconductors at infrared and THz frequencies~\cite{west2010}, and wire-mesh metamaterials in the microwave regime~\cite{pendry1996}. We envisage that the conceptual framework outlined here may lead to a new class of metamaterial designs exploiting the properties of metals near the plasma frequency.

This work is part of the research program of FOM, which is financially supported by NWO. It is supported by the Joint Solar Programme (JSP) of FOM, which is co-financed by NWO and Stich\-ting Shell Research.


\clearpage

\section{Supplementary information}

\subsection{Modal symmetry and the sign of the coupling constant}

As argued in the main text, whether energy is refracted positively or negatively is controlled by the sign of the coupling constant $\kappa$ and the mode wavevector $\beta$. The coupling constant is approximated in the weak coupling limit by Eq.~(3). A more general expression for the coupling strength between a waveguide aligned along $\mathbf{\hat{x}}$ with dielectric constant profile $\varepsilon\left(z\right)$ and an identical adjacent waveguide spaced at a distance $d$ is~\cite{marcuseB}
\begin{equation}
\kappa = \omega \varepsilon_0 \frac{\int_{-\infty}^{\infty} \left[\varepsilon\left(z\right)-\varepsilon\left(z-d\right)\right] \mathbf{E}^{\dag}\left(z\right) \cdot \mathbf{E}\left(z-d\right) dz}{\int_{-\infty}^{\infty} \mathbf{\hat{x}} \cdot \left( \mathbf{E}^{\dag} \times \mathbf{H} + \mathbf{E} \times \mathbf{H}^{\dag} \right) dz},
\label{eqgeneralkappa}\tag{S1}
\end{equation}
where $\mathbf{E}$ and $\mathbf{H}$ are the mode fields of a single waveguide. The use of the adjoint field instead of the complex conjugate allows the dielectric constants to be complex~\cite{marcuseB}. The fact that the integrand in the numerator scales with $\left[\varepsilon\left(z\right)-\varepsilon\left(z-d\right)\right]$ can be understood when relating the coupling constant to the perturbation of the field in one waveguide by the presence of the adjacent waveguide, with the difference of the dielectric constant profiles taking the role of a potential difference. In the MDM waveguides considered in this work, this difference vanishes everywhere except in the dielectric regions, and Eq.~(\ref{eqgeneralkappa}) reduces to Eq.~(3). Apart from brute-force evaluation of Eq.~(3), the \emph{sign} of $\kappa$ can also be found from simple arguments. In the case of lossless media, the adjoint field $\mathbf{E}^{\dag}$ could be replaced by the complex conjugate $\mathbf{E^*}$ and the denominator of Eq.~(3) then equals four times the total power carried by the waveguide, which is evidently a positive quantity. The factor $\left(\varepsilon_\mathrm{d}-\varepsilon_\mathrm{m}\right)$ is positive as well. Therefore, the sign of $\kappa$ is determined by the overlap integral of the modal electric field $\mathbf{E}\left(z\right)$ with the electric field in a neighboring waveguide $\mathbf{E}\left(z-d\right)$, evaluated over the dielectric region.

\begin{figure}[]
\includegraphics[width=7cm]{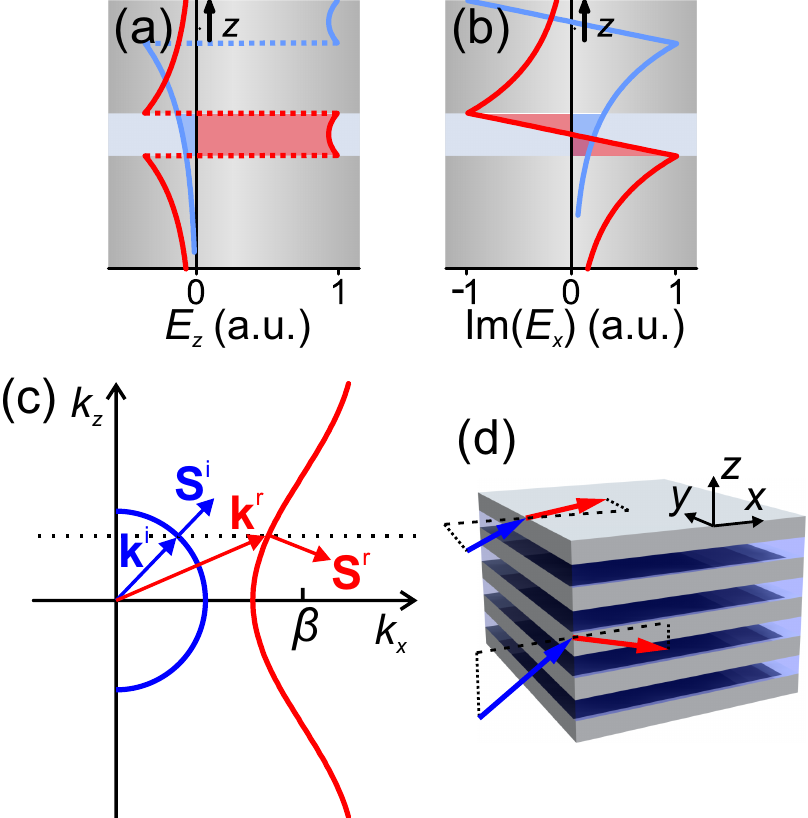}
\caption{(Supplemental). Coupling and refraction of the symmetric mode in the positive-index regime. (a) Typical $E_z\left(z\right)$ mode profile of the symmetric mode in a MDM waveguide for $\omega<\omega_\mathrm{sp}$ (red), and the displaced field $E_z\left(z-d\right)$ (blue), showing how the mode overlap in the dielectric core is negative. (b) $E_x\left(z\right)$ (red) and $E_x\left(z-d\right)$ (blue) mode profile, with also negative overlap. (c) Sketch of the wavevector diagram with the isofrequency contour of a homogeneous medium from which a plane wave with wavevector $\mathbf{k}^\mathrm{i}$ is incident (blue), and the isofrequency contour of the symmetric mode in a coupled stack of MDM waveguides, with $\beta>0$ and $\kappa>0$ (red). (d) Refraction of energy is negative in the $xz$-plane, but positive in the $xy$-plane.}
\label{FigS1}
\end{figure}

As an example, we first consider the symmetric mode in a MDM waveguide at frequencies $\omega<\omega_\mathrm{sp}$, for which both $\beta$ and $v_\mathrm{g}$ are positive. As sketched in Fig.~\ref{FigS1}~(Supplemental)(a), the overlap of the $E_z\left(z\right)$ component in the dielectric core with the field $E_z\left(z-d\right)$ in the cladding of the adjacent waveguide ($\int_d E_z^\dag\left(z\right) E_z\left(z-d\right) dz$) is negative, since $E_z$ switches sign across a metal-dielectric boundary.
We used the symmetry of both $E_z$ and $H_y$ to label this mode as being `symmetric'. In contrast, the $z$ profile of the longitudinal component $E_x$ of a symmetric mode is itself antisymmetric (and vice versa). However, unlike $E_z$, $E_x$ does not change sign across a metal-dielectric interface, and hence the net overlap integral $\int_d E_x^\dag\left(z\right) E_x\left(z-d\right) dz<0$. This can be judged from the mode profile of $E_x$ in Fig.~\ref{FigS1}~(Supplemental)(b). Note that we plot $\Im{E_x}$ since $E_x\left(x,z\right)$ is imaginary for the particular value of $x$ that was used to calculate the (real) $E_z$ profile in Fig.~\ref{FigS1}~(Supplemental)(a). Since the contributions of $E_z$ and $E_x$ to the overlap integral in Eq.~(3) are both negative, $\kappa$ is always negative. As a result, the isofrequency contour (Eq.~(2)) in the $xz$-plane of a stack of MDM waveguides for $\omega<\omega_\mathrm{sp}$ is such that energy is refracted negatively~\cite{fan2006aB}, as can be seen from the wavevector diagram in Fig.~\ref{FigS1}~(Supplemental)(c). The isofrequency contour resembles hyperbolic dispersion for small $k_z$, and the arguments above could be similarly applied to explain negative refraction in other indefinite media~\cite{yao2008B}. However, refraction of the wavevector is clearly positive and angle-dependent in this right-handed metamaterial, and refraction of both phase and energy in the $xy$-plane is exclusively positive as well, as sketched in Fig.~\ref{FigS1}~(Supplemental)(d)~\cite{podolskiy2007}.

\begin{figure}[]
\includegraphics[width=5cm]{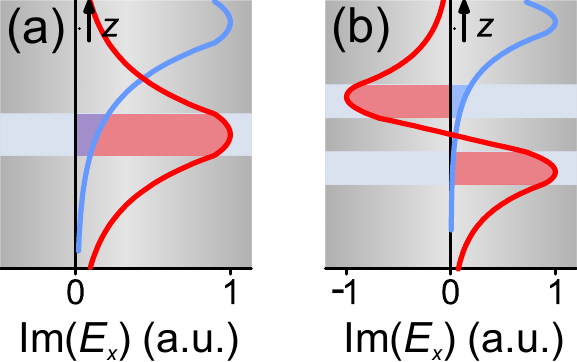}
\caption{(Supplemental). (a) Field profile of longitudinal electric field $E_x\left(z\right)$ of the antisymmetric mode in a MDM waveguide (red), as well as the displaced field $E_x\left(z-d\right)$ (blue), having positive overlap. (b) Profiles of $E_x\left(z\right)$ (red) and $E_x\left(z-d\right)$ (blue) of the symmetric mode in a pair of waveguides in the backwards wave regime, with a negative overlap.}
\label{FigS2}
\end{figure}

The antisymmetric backwards mode in a thin MDM waveguide for $\omega>\omega_\mathrm{sp}$ exhibits opposite behavior compared to the symmetric mode in many respects. The field $E_z\left(z-d\right)$ in the cladding of the displaced waveguide now has a net positive overlap with the field $E_z\left(z\right)$ in the core of the waveguide due to the antisymmetric nature of this field component, as was clear from Fig.~2(b). Of course, the longitudinal component $E_x$ should be analyzed in a similar fashion. In fact, for the backwards waves considered here the longitudinal component generally constitutes the largest contribution to the overlap integral, especially close to the cutoff frequency. Its profile $E_x\left(z\right)$ is sketched in Fig.~\ref{FigS2}~(Supplemental)(a).
The $E_x$ field of an antisymmetric mode is itself symmetric, and it is continuous at the metal-dielectric interfaces. As a result, $\int_d E_x^\dag\left(z\right) E_x\left(z-d\right) dz>0$, just like the overlap integral of $E_z$, and therefore we know that $\kappa$ must be positive.

Finally, the $E_x$ field profile of the symmetric mode in a pair of waveguides (i.e., the odd superposition of the antisymmetric modes of the individual MDM waveguides) is depicted in Fig.~\ref{FigS2}~(Supplemental)(b). Since the symmetry is opposite to that of the backwards wave in a single MDM waveguide, in this case $\kappa<0$. This now fully explains why refraction of both phase and energy are negative in a metal-dielectric multilayer stack with alternating thick and thin metal films.

The above analysis has again illustrated how mode symmetry governs the sign of the coupling constant. In particular, in metal-dielectric waveguides (where the transverse component $E_z$ switches sign at the core-cladding interface), symmetric modes yield $\kappa<0$ and antisymmetric modes yield $\kappa>0$. We note that in all-dielectric waveguides --- in which $E_z$ does not switch sign across an interface --- the relationship between symmetry and the sign of the coupling constant will in most cases be reversed. The connection is however less general in that case, since the transverse and longitudinal components have opposing contributions to the overlap integral.

\subsection{Isofrequency contours in the presence of absorption}

\begin{figure}[t]
\includegraphics[width=8.6cm]{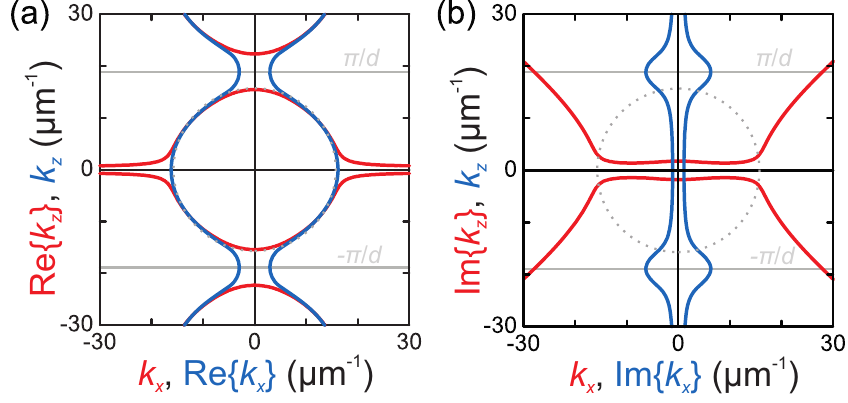}
\caption{(Supplemental). Calculated wavevector diagrams for the example structure described in the main text, including absorption in the Ag, taking either $k_x$ or $k_z$ to be complex. (a) $\Re{k_x}$ as a function of real $k_z$ (blue) and $\Re{k_z}$ as a function of real $k_x$ (red). (b) $\Im{k_x}$ as a function of real $k_z$ (blue) and $\Im{k_z}$ as a function of real $k_x$ (red). The grey dotted curves represent $k_x^2+k_z^2=k_0^2$.} \label{FigS3}
\end{figure}

It is important to evaluate the effect of realistic absorption losses and compare the results to the lossless case. We use $\varepsilon_\mathrm{m}=-4.43+0.21i$ in the example described in the main text at $\lambda_0=400$~nm~\cite{johnson1972B}, to plot the wavevector diagrams in Fig.~\ref{FigS3}(a) for two cases; taking $k_z$ real and $k_x$ complex (blue curve, relevant in the context of propagation and damping along the layers, and refraction at an interface of the metamaterial parallel to the $yz$-plane) and taking $k_x$ real and $k_z$ complex (red curve, useful in the context of propagation normal to the layers and refraction at an interface parallel to the $xy$-plane). Because the red curve still follows the $n=-1$ contour (the grey dotted curve) at small $\left|k_x\right|$ and the blue curve does the same at small $\left|k_z\right|$, it can be concluded that losses do not significantly affect the value of the effective index, nor its isotropic character. In Fig.~\ref{FigS3}~(Supplemental)(b), the imaginary part of the wavevectors is plotted for both cases.
It shows that $\Im{k_z}$ is relatively small for $-k_0<k_x<k_0$, where the mode has a propagating character. Outside this regime, the mode is largely evanescent. Likewise, $\Im{k_x}$ is small for $-k_0<k_z<k_0$, and has a maximum value at $k_z=\pi/d$. The figure of merit $\mathrm{FOM}=\Re{\left|\mathbf{k}\right|}/\Im{\left|\mathbf{k}\right|}$ is only unambiguously defined along the $x$ and $z$ axes, since in those directions the symmetry of the geometry dictates that $\mathbf{k}$ and $\mathbf{S}$ are entirely parallel or antiparallel. The FOM varies from 14.7 along the $x$ direction (obtained by taking $k_z=0$ and $k_x$ complex) to 8.7 along $\mathbf{\hat{z}}$ (with $k_x=0$ and $k_z$ complex).

\subsection{Calculation of coupling efficiency}

The coupling efficiency of a wave incident from air on an interface of the metamaterial parallel to the $yz$ plane can be evaluated in a finite-difference time-domain (FDTD) simulation by integrating the power flow inside the metamaterial at a depth of 1~$\mu$m below the interface. At this depth, only the propagating backwards mode exists in the waveguide. Taking into account the known propagation losses over the first 1~$\mu$m, the fraction of energy coupled to the mode of interest is calculated.

By varying the layer thicknesses in the array of coupled waveguide pairs, in particular the relative thicknesses of the two metal films, the coupling efficiency can be controlled to some extent. For example, at a free-space wavelength of 474~nm, a coupling efficiency of 19.6\% was found for a stack consisting of Ag (using permittivity from ref.~\cite{johnson1972B}) and a dielectric with a refractive index of 3.12. The dielectric thickness was 44~nm, and the thicknesses of the alternating metal films were 24 and 90~nm, respectively, resulting in an effective index of $-2.1$ at normal incidence. We note that we limited the investigated structures to those consisting of only one type of metal and one type of dielectric, and we have chosen the thickness of the two dielectric films in the four-layer unit cell to be equal. Including different materials and allowing the thicknesses of the two dielectric layers to be different would introduce additional degrees of freedom to further optimize impedance matching. Moreover, it may be possible to improve impedance matching by properly structuring the interface of the metamaterial~\cite{russel1995}.


\begin{thebibliography}{34}
\bibitem[{foo()}]{footnotenimcoupling2}
\bibinfo{note}{We note that the even superposition has decreasing
  $\left|\beta\right|$ with decreasing distance between the waveguides in the
  pair, and is in fact often cut-off.}
\end{thebibliography}

\end{document}